\documentclass{aa}  

\usepackage{graphicx}
\usepackage{txfonts}
\usepackage[colorlinks=true, citecolor=blue, urlcolor=blue]{hyperref}
\usepackage{placeins}
\usepackage{color, soul}
\usepackage{ulem}
\usepackage{natbib}
\usepackage{lipsum}
\bibpunct{(}{)}{;}{a}{}{,} 
\usepackage{tabularx}  

\begin{document}

   \title{On the orbital variability of gamma-ray emission in $\gamma^2$ Velorum}

   \titlerunning{The emission from $\gamma^2$ velorum}

   \author{Agnibha De Sarkar
          }

   \institute{Institute of Space Sciences (ICE, CSIC), Campus UAB, Carrer de Can Magrans s/n, E-08193 Barcelona, Spain \\
             \email{desarkar@ice.csic.es}\\
            }
    
   \date{Received XXXX; accepted YYYY}

 
\abstract
   {Colliding wind binaries (CWBs) are promising sources of high-energy gamma-ray emission driven by shock acceleration of particles at wind interaction zones. The nearby CWB system $\gamma^2$~Velorum (WR 11), composed of a Wolf–Rayet (WR) and an O-star, has been recently associated with GeV gamma-ray emission observed by \textit{Fermi}-LAT, including showing evidence of orbital variability. This offers a valuable opportunity to test models of phase-dependent hadronic emission and absorption in CWBs.}
   {We aim to explain both the spectral energy distribution (SED) and orbital variability of gamma-ray emission from $\gamma^2$~Velorum using a physically motivated phase-dependent hadronic model.}
   {We consider the injection of accelerated relativistic protons based on the WR wind's kinetic energy intercepted at the wind collision region (WCR), and calculate the resulting phase-dependent hadronic gamma-ray emission assuming a proton conversion efficiency $\eta_p$ and accounting for energy-dependent diffusion, advection, conical shock interception and the evolution of the effective acceleration volume, assumed to scale with the WCR, with the orbital phase. Gamma-ray emission from hadronic interactions is attenuated by $\gamma$--$\gamma$ absorption, calculated via full angular integration over both stellar photon fields.}
   {Our model, including the attenuation resulting from the $\gamma$--$\gamma$ absorption, successfully reproduces the observed SED and is consistent with the apastron-to-periastron flux ratio, resulting in a dip in emission at periastron passage, while an increase occurs during the apastron.}
   {Our findings support the conclusion that the observed orbital modulation is primarily driven by geometric variations of the WCR. This underscores the significant influence of evolving orbital geometry on the high-energy gamma-ray light curves of $\gamma^2$ Velorum. As CWBs emerge as a potential new class of high-energy gamma-ray sources, advancing our understanding will require more detailed magnetohydrodynamic (MHD) modeling of the wind interaction dynamics in these systems.}

   {}

   \maketitle
%

\section{Introduction}

Non-thermal emission from gamma-ray binaries—systems composed of a massive star and a compact object, typically a pulsar or black hole—has been robustly detected at high energies (HE; GeV) and very high energies (VHE; TeV) (see for e.g., \citealt{dubus06, dubus13, maraschi81, huber21, bogovalov08, bednarek09, boschramon12, hinton09, desarkar22}). In addition, the recurrent nova binary system RS Ophiuchi, comprising of a white dwarf and a red giant companion, has recently been observed in both GeV and TeV gamma rays (see for e.g., \citealt{tatischeff07,acciari22, hess22, cheung22, diesing23, desarkar23}), further indicating the presence of efficient particle acceleration. A distinct subclass of binary systems, colliding wind binaries (CWBs), consists of two massive stars whose powerful stellar winds interact, forming a wind collision region (WCR) capable of accelerating particles to relativistic energies (see for e.g., \citealt{eichler93, dougherty2000, debecker13, benaglia01, benaglia03, reimer06, reitberger14, grimaldo19}). These systems are promising candidates for gamma-ray emission via hadronic and/or leptonic processes.

The only firmly established GeV-emitting CWB to date is $\eta$ Carinae (see for e.g., \citealt{farnier11, bednarek11, ohm15, gupta17, reitberger15, balbo17}), which has also been detected at TeV energies \citep{hess20}. Another notable system is $\gamma^2$ Velorum (WR 11), the closest known CWB at a distance of $336^{+8}_{-7}$ pc \citep{north07}, consisting of a WC8 star and an O7.5 companion \citep{henley05} in a $78.53 \pm 0.01$ day orbit \citep{schmutz97} with a moderate eccentricity of $0.334 \pm 0.003$ \citep{north07}. This system exhibits both thermal \citep{purton82, benaglia19} and non-thermal \citep{chapman99, benaglia16} radio emission, explained by synchrotron emission of accelerated electrons at the WCR. \cite{debecker13} explains the non-thermal radio emission, including its absorption in the WR star wind. The WCR has been observed in ultraviolet \citep{stlouis93} and X-rays by ROSAT \citep{willis95}, ASCA \citep{rauw2000}, Chandra \citep{skinner01}, and XMM-Newton \citep{schild04}, including a wide shock-cone opening half angle of approximately $\approx 85^\circ$ as revealed by Chandra observations \citep{henley05}.

While $\gamma^2$ Velorum has been previously associated with a gamma-ray source \citep{pshirkov16}, only recent \textit{Fermi}-LAT data analysis has revealed a significant high-energy source 4FGL J0809.5-4714 spatially consistent with $\gamma^2$ Velorum, with hints of orbital variability \citep{devesa20}, where flux increases around apastron and decreases near periastron. This behavior differs from $\eta$ Carinae, where maximum $\gamma$-ray emission occurs near periastron and vice versa. This phenomenon thus motivates a detailed modeling of the phase-resolved emission.

Although a leptonic scenario has been proposed to explain the gamma-ray emission from $\gamma^2$ Velorum \citep{tatischeff04}, comparatively more recent literature favors a hadronic origin, with the emission primarily arising from neutral pion ($\pi^0$) decay following proton acceleration in the WCR \citep{reitberger17}. However, these earlier studies did not particularly focus on the orbital modulation of the gamma-ray emission, which has only been identified relatively recently. In this paper, we present a simple semianalytic treatment of the phase-dependent hadronic emission for $\gamma^2$ Velorum, incorporating both the orbital geometry and $\gamma$–$\gamma$ absorption. We demonstrate that reproducing the observed modulation indeed necessitates an intrinsic geometric variation of the acceleration region, while attenuation effects along the line of sight play a comparatively minor role.

In Sect. \ref{model}, we outline the theoretical framework adopted in this study. Subsect. \ref{env} describes the physical environment of the $\gamma^2$ Velorum system. The proton injection model is detailed in Subsect. \ref{proton}, while the treatment of $\gamma$--$\gamma$ absorption is presented in Subsect. \ref{gamma}. The results of our phase-resolved modeling are discussed in Sect. \ref{results}, followed by our main conclusions in Sect. \ref{conclusion}.
\section{The model}\label{model}

\subsection{CWB environment}\label{env}

In CWBs such as $\gamma^2$~Velorum, each massive star drives a powerful, accelerated stellar wind characterized by high mass-loss rates and supersonic terminal velocities. As these winds expand outward, they eventually collide, forming a wind interaction zone bounded by two strong shocks and a contact discontinuity (CD) that separates the shocked plasma of each star. This structure, i.e., the WCR, is shaped by the wind momentum balance and orbital geometry, and typically adopts a curved morphology that wraps around the star with the weaker wind momentum flux—namely, the O-star companion in the case of $\gamma^2$~Velorum. Across these shocks, the initially supersonic winds are abruptly decelerated and compressed, resulting in the formation of high-temperature plasma and an enhancement of the local magnetic field. These shocked regions are ideal sites for diffusive shock acceleration (DSA), wherein charged particles can gain energy through repeated crossings of the shock front. Both electrons and protons can be accelerated to relativistic energies in this environment, producing observable non-thermal emission via leptonic and hadronic processes, respectively. The efficiency of acceleration and the resulting emission depend sensitively on the local physical conditions—such as the magnetic field strength, ambient density, and orbital geometry—which vary with orbital phase due to the system’s eccentricity.

\begin{table*}[t]
\centering
\caption{Stellar and wind parameters of the O-star and WR-star components of the $\gamma^2$~Velorum system.}
\label{tab1}
\begin{tabularx}{\textwidth}{lXX}
\hline\hline
Parameter 
& O-star (O7.5) 
& WR star (WC8) \\
\hline
Mass $M_*$ [$M_\odot$] 
& 28.5~[1,2] 
& 9.0~[2] \\
Radius $R_*$ [$R_\odot$] 
& 17.0~[2] 
& 6.0~[2] \\
Effective temperature $T_*$ [K] 
& 35000~[1] 
& 56000~[2] \\
Luminosity $L_*$ [$L_\odot$] 
& $2.8 \times 10^5$~[2] 
& $1.7 \times 10^5$~[2] \\
Mass-loss rate $\dot{M}_*$ [$M_\odot\,\mathrm{yr}^{-1}$] 
& $1.78 \times 10^{-7}$~[1] 
& $8.0 \times 10^{-6}$~[2,3] \\
Terminal velocity $V_{*, \infty}$ [km\,s$^{-1}$] 
& 2500~[1] 
& 1450~[2,3] \\
Surface magnetic field $B_*$ [G] 
& 100~[4] 
& 100~[4] \\
\hline
\end{tabularx}

\vspace{0.5em}
\begin{minipage}{\textwidth}
\footnotesize
\textbf{Note:} The references are taken from   
[1] \cite{demarco00},  
[2] \cite{north07},  
[3] \cite{schild04},  
[4] \cite{reitberger17}.
`` * '' is replaced by O for O-star, and W for WR star in this work.
\end{minipage}
\end{table*}

\begin{table}[t]
\centering
\caption{Orbital parameters of the $\gamma^2$~Velorum system used in this study.}
\label{tab2}
\begin{tabularx}{\columnwidth}{lX}
\hline\hline
Parameter & Value \\
\hline
Distance $D$ [pc] 
& 336~[1] \\
Orbital separation $a$ [$R_\odot$] 
& 258~[1] \\
Orbital period $P$ [days] 
& 78.53~[2] \\
Eccentricity $e$ 
& 0.334~[2] \\
Inclination $i$ [deg] 
& 65~[2] \\
\hline
\end{tabularx}

\vspace{0.5em}
\begin{minipage}{\columnwidth}
\footnotesize
\textbf{Note:} The references are taken from    
[1] \cite{north07},  
[2] \cite{schmutz97}.
\end{minipage}
\end{table}

As discussed earlier, a wide shock-cone opening half angle of approximately $85^\circ$ was observed in $\gamma^2$~Velorum \citep{henley05}, which can be attributed to the effects of radiative braking \citep{reitberger17}. 
{The sudden radiative braking effect in CWBs, especially in WR + O type binaries, has been previously extensively studied both analytically and numerically in \cite{gayley97}.
The authors have shown that radiative braking may prove to be a necessary effect, as in some systems, not considering this effect may render the hydrodynamical wind-wind ram balance impossible.
}
In this scenario, the intense radiation field of the O-star decelerates the approaching WR wind before the collision, effectively reducing its velocity significantly from the terminal value near the WCR. This radiative braking alters the local momentum balance and leads to a wider shock cone than that would be expected from unimpeded wind interactions alone. Motivated by this, we adopt an effective WR wind velocity that is a reduced fraction of its nominal terminal value. This choice not only reproduces the observed shock-cone opening half angle, but also provides a more realistic estimate of the wind momentum flux in the vicinity of the collision region. 

The key physical parameters for the two companion stars of the $\gamma^2$~Velorum system used in this study are summarized in Table~\ref{tab1}. The orbital parameters of the source are given in Table \ref{tab2}.  Based on these parameters, we first estimate the wind momentum balance ratio,
\begin{equation}\label{eq1}
\eta_\phi = \frac{\dot{M}_{O} V_{O, \infty}}{\dot{M}_\mathrm{W} V_\mathrm{W, brak}},
\end{equation}
where $\dot{M}_{\mathrm{W}}$ and $\dot{M}_{\mathrm{O}}$ are the mass-loss rates of the WR and O-stars, respectively. The mass-loss rate of the WR star in $\gamma^2$~Velorum remains uncertain, as different observational methods yield significantly different values. Polarimetric analyses suggest a rate of approximately $8.0 \times 10^{-6} M_\odot \mathrm{yr}^{-1}$, while estimates based on radio emission indicate a higher value of $\sim 3.0 \times 10^{-5} M_\odot \mathrm{yr}^{-1}$. This discrepancy is likely a consequence of wind clumping, which can significantly enhance the free-free radio emission and thereby inflate mass-loss estimates derived from radio observations. In contrast, polarimetric measurements are less sensitive to small-scale density inhomogeneities and are generally considered more robust in systems where clumping is expected. For these reasons, and to avoid overestimating the wind momentum flux of the WR star, we adopt the polarimetric value in this study.

Next, $V_{O, \infty}$ is the O-star wind terminal velocity, whereas $V_\mathrm{W, brak}$ is the radiatively braked velocity of the WR wind, obtained by considering a fraction of the WR wind terminal velocity $V_{W, \infty}$. To obtain this fraction, we solve the shock-cone opening half angle equation from \cite{eichler93}, i.e.,

\begin{equation}\label{eq2}
    \theta \ { \mathrm{(in \ deg)}} \approx 2.1 \left(1 - \frac{\eta_{\phi}^{2/5}}{4}\right) \eta_{\phi}^{1/3} { \left(\frac{180^{\circ}}{\pi}\right)}, \: \hspace{0.5cm} \mathrm{for\: 10^{-4} \: \leq \: \eta_{\phi} \: \leq \: 1 },
\end{equation}
for $\theta \approx 85^{\circ}$. We find that the wide shock-cone opening half-angle can be reconciled for the braked WR wind velocity being around 5$\%$ of its nominal terminal velocity at the collision region. 
{ The braking radius, where wind braking becomes significant, can be calculated from the analytical formulation provided in \cite{gayley97}. We found that for the reduction in WR wind velocity due to radiative braking, the braking radius falls very close to the stagnation point radius, which is expected in this scenario. 
We further found that without the inclusion of radiative braking, the ram balance could not be achieved during the periastron, as was resonated in \cite{gayley97}.
We find the reflection factor $S$ (see \cite{gayley97}), representing the radiative momentum from the O-star, proves to be significant in this case, further ensuring that sudden braking of WR wind due to the O-star radiative momentum is highly expected in $\gamma^2$ Velorum. 
}
We have considered the braked WR wind velocity $V_\mathrm{W, brak}$ for further calculation.

The shock apex, or the stagnation point, defined as the location along the line connecting the two stars where the ram pressures of the winds equilibrate, lies closer to the O-star for the case of $\gamma^2$ Velorum. The distance of the shock from each star—and hence the overall shape of the WCR—can be derived from the aforementioned ratio. In an elliptical orbit, the separation between two stars can be given by,

\begin{equation}
    R_{\rm sep} =  a (1 - e^2)/(1+ e\:cos(\phi)).
\end{equation}

Using this equation, and assuming the collision of two spherical winds, one can calculate the distances $R_{\rm sh, O}$ and $R_{\rm sh, W}$ from the O- and WR stars to the wind interaction region along the line joining the O- and WR stars following,

\begin{equation}
    R_{\rm sh, O} = \frac{\sqrt{\eta_{\phi}}}{1 + \sqrt{\eta_{\phi}}} R_{\rm sep} \hspace{0.3cm} \mathrm{and} \hspace{0.3cm} R_{\rm sh, W} = \frac{1}{1 + \sqrt{\eta_{\phi}}} R_{\rm sep}.
\end{equation}
Furthermore, due to the eccentric orbit of $\gamma^2$~Velorum, the stellar separation—and thus the location of the WCR—varies with orbital phase. This variation leads to a changing shock geometry, impacting the local physical conditions such as density, and the apex magnetic field strength. To properly account for these phase-dependent effects, we calculate the local wind velocity of the O-star at the position of the WCR using a $\beta$-velocity law. 
We adopt an index of $\beta_{\rm O} = 0.8$, as appropriate for O-stars \citep{gayley97}. Given the O-star radius $R_\mathrm{O}$, the local wind velocity at the standoff distance $R_{\rm sh, O}$ is computed as,
\begin{equation}
V_{\rm O, local} = V_{\rm O, \infty} \left(1 - \frac{R_{\rm O}}{R_{\rm sh, O}} \right)^{\beta_{\rm O}}.
\end{equation}
{For the WR star, we adopt a constant local wind velocity, \( V_{\rm W, local} = V_{\rm W, brak} \), corresponding to the effective braked velocity. This choice is motivated by the assumption that, due to a much smaller radius of the WR star, its wind acceleration takes place in a region largely unaffected by the influence of the companion O-star \citep{gayley97}. It is important to note that radiative braking acts predominantly along the line connecting the two stars, where the winds directly interact with each other.
}

These local wind velocities are used to derive the magnetic field strength from each star at the WCR apex and the WR wind number density. The kinetic luminosity of the WR wind entering the shock, as well as the advection timescale, is also estimated using the local velocities.  

To compute the magnetic field strength at the location of the WCR, we adopt a structured magnetic field prescription following the approach of \citet{eichler93}, in which the field transitions from a dipole-dominated regime near the stellar surface to a toroidal configuration at larger radii.
We first calculate the Alfvén radius by defining the wind magnetic confinement parameter \( \eta_* \) as \citep{uddoula02}:
\begin{equation}
\eta_* = \frac{B_*^2 R_*^2}{\dot{M}_* V_{*, \mathrm{local}}},
\end{equation}
where \( B_* \) is the surface magnetic field strength, \( R_* \) is the stellar radius, \( \dot{M}_* \) is the mass-loss rate, and \( V_{*, \mathrm{local}} \) is the local wind velocity (or the braked velocity in case of the WR). The Alfvén radius \( R_{\mathrm{A}} \) is estimated using the strong confinement scaling \citep{uddoula08}:
\begin{equation}
R_{\mathrm{A}} \approx R_* \left( 0.3 + \eta_*^{1/4} \right).
\end{equation}
Given the stellar surface rotational velocity \( V_{\mathrm{rot}}\:(\sim 0.2 V_{\mathrm{*, \infty}}) \), the magnetic field strength \( B(r) \) at a radial distance \( r \) from the center of the star is then defined in three regimes:

\begin{itemize}
    \item Dipole regime (\(R_* \leq r < R_{\mathrm{A}} \)):
    \begin{equation}
    B(r) = B_* \left( \frac{R_*}{r} \right)^3,
    \end{equation}
    
    \item Radial regime (\( R_{\mathrm{A}} \leq r < R_* \left( \frac{V_{*, \mathrm{local}}}{V_{\mathrm{rot}}} \right) \)):
    \begin{equation}
    B(r) = B_* \frac{R_*^3}{R_{\mathrm{A}} \, r^2},
    \end{equation}
    
    \item Toroidal regime (\( r \geq R_* \left( \frac{V_{*, \mathrm{local}}}{V_{\mathrm{rot}}} \right) \)):
    \begin{equation}
    B(r) = B_* \left( \frac{V_{\mathrm{rot}}}{V_{*, \mathrm{local}}} \right) \frac{R_*^2}{R_{\mathrm{A}} \, r}.
    \end{equation}
\end{itemize}

This three-zone prescription is applied separately to each star in the binary. The magnetic field at the WCR location is calculated considering the contributions from both stars.

Note that the shock can undergo a strong MHD compression in the region. 
{The value of the corresponding compression ratio depends on whether the shock is in a strong adiabatic regime or has transitioned into the radiative regime.}
For a strong, adiabatic shock with adiabatic index \( \gamma = 5/3 \), the compression ratio is given by $\chi \approx 4$. 
{However, if the shock has transitioned into the radiative regime, then the compression ratio can increase way above the value of $\chi \approx 4$.}
{To determine the thermal regime of the shock, we adopt the diagnostic criterion outlined in \citet{stevens92} and \citet{pittard10}, which introduces the parameter \( \xi \), that can be calculated by,

\begin{equation}
    \xi \approx \frac{(V_{*, \mathrm{local}} / 10^8~\mathrm{cm\,s^{-1})}^4 (R_{\mathrm{sh,*}} / 10^{12}~\mathrm{cm})}{(\dot{M}_{\mathrm{*}} / 10^{-7}~M_\odot\,\mathrm{yr}^{-1})}.
\end{equation}
The parameter $\xi$ acts as a characteristic measure for the importance of cooling. If $\xi \gtrsim 1$, then the wind can be assumed to be adiabatic, whereas if $\xi << 1$, the wind can be assumed to be in the radiative regime.
Using the parameters of $\gamma^2$~Velorum, we find that while the O-star shock remains in the adiabatic regime ($\xi > 1$) across all orbital phases, the WR shock consistently transitions into the radiative regime, with $\xi << 1$ at all orbital phases.
A radiative WR shock indicates a stronger post-shock compression and amplified magnetic field. In contrast, the adiabatic nature of the O-star shock leads to weaker compression and consequently less efficient magnetic field amplification.
To incorporate this into our modeling, we adopt a higher compression ratio for the WR wind, $\chi_\mathrm{W} \approx 10$, consistent with its radiative nature. For the O-star wind, we retain the standard adiabatic compression ratio, $\chi_\mathrm{O} \approx 4$.

}
Assuming both shocks compress the upstream magnetic field, the total post-shock magnetic field at the WCR apex is given by:
\begin{equation}
B_{\mathrm{apex}} = \left[ \left( B_{\mathrm{sh,O}} \cdot \chi_O^{2/3} \right)^2 + \left( B_{\mathrm{sh,W}} \cdot \chi_W^{2/3} \right)^2 \right]^{1/2}.
\end{equation}
This combined magnetic field strength \( B_{\mathrm{apex}} \) is then used in calculating the maximum particle energies in the WCR.

In our model, we consider relativistic protons interacting primarily with the shocked WR wind material at the WCR. 
{Crucially, given the presence of WC8 subtype star in $\gamma^2$ Velorum, WR wind is hydrogen-poor, but primarily composed of heavier elements such as Helium (He), Carbon (C), and Oxygen (O), as is observed from their strong line emissions (see \cite{crowther07} and references therein). }
The pre-shock target material number density at a radial distance \( R_{\mathrm{sh,W}} \) from the WR star, assuming a steady, spherically symmetric wind, is given by the standard continuity equation: \( n_{\mathrm{w}} = \dot{M}_{\mathrm{W}} / (4\pi R_{\mathrm{sh,W}}^2 V_{\mathrm{W,local}} {\bf \mu} m_{\mathrm{p}}) \), where \( \dot{M}_{\mathrm{W}} \) and \( V_{\mathrm{W,local}} \) are the same parameters discussed above, {$\mu$ is a factor signifying the mean molecular weight} and \( m_{\mathrm{p}} \) is the proton mass. {However, since the gamma-ray producing protons are assumed to interact in the post-shock region on the side of the WR wind, we adopt a shocked material density enhanced by a compression factor $\chi_W (\approx 10)$, as expected for the radiative nature of the WR wind shock.} This yields a post-shock density of
\begin{equation}
n_{\mathrm{w,shocked}} = \chi_W \cdot n_{\mathrm{w}} = \frac{\chi_W \dot{M}_{\mathrm{W}}}{4 \pi R_{\mathrm{sh,W}}^2 V_{\mathrm{W,local}} \mu m_{\mathrm{p}}},
\end{equation}
which we use as the effective target density for the hadronic interactions. 
{For a representative composition of 55$\%$ He, 40$\%$ C, and 5$\%$ O by mass (see, for e.g., \cite{crowther07}), we derive the value of $\mu \approx$ 5.75, which we adopt in computing the number density of targets for hadronic interactions.}
This quantity plays a key role in setting the normalization of the hadronic gamma-ray emission, as it directly governs the pion production rate at the WCR apex.

\subsection{Proton injection}
\label{proton}

To model the hadronic gamma-ray emission from $\gamma^2$~Velorum, we compute the phase-dependent injection of relativistic protons at the WCR. This injection process is governed by a combination of physical quantities: the maximum energy attainable by shock-accelerated protons; the WR wind kinetic energy available for particle acceleration over the advection timescale; the fraction of the WR wind intercepted by the WCR; the fraction of accelerated protons retained in the emission region before they can escape; and the effective volume of the acceleration region, which controls the number of particles that can be accelerated. Most of these quantities are modulated by the binary's orbital geometry and wind interaction dynamics, and together they determine both the normalization and shape of the proton injection spectrum at each phase.
\noindent
{Despite the hydrogen-poor nature of the WR wind, the presence of even a trace population of protons can suffice for DSA. Additionally, protons can be supplied at the WCR by the hydrogen-rich O-star wind through mixing or shock interface instabilities. For simplicity, we assume that the accelerated population is predominantly composed of protons, consistent with standard cosmic-ray acceleration theory.}
{The target material for hadronic interactions is modeled using the WR wind composition. To account for the heavy-element-enriched nature of the WR wind in $\gamma^2$~Velorum, we adopt a nuclear enhancement factor of $\epsilon_{\mathrm{enh}} \approx 2$ in our hadronic gamma-ray calculations. This accounts for the increased pion production efficiency in proton interactions with heavier nuclei (e.g., \citealt{kafe14}), and is appropriate for the energy range considered ($\gtrsim 1$~GeV).}

The maximum energy cutoff of the proton distribution, \( E_{p,\mathrm{max}} \), is set by comparing the characteristic acceleration time with energy loss and escape timescales. Following the formalism in \citet{bednarek11}, we consider two distinct regimes that limit acceleration: the loss-limited regime, where losses due to hadronic interaction dominate, and the escape-limited regime, where particles advect out of the WCR before significant energy losses occur.

In the loss-limited scenario, the maximum energy is given by:
\begin{equation}
E_{p,\mathrm{max}} \approx 6.3\,\xi_{-5}\,B_{\mathrm{apex}}\,R_{13}^2\,v_3 / \dot{M}_{-4} \quad \mathrm{TeV},
\end{equation}
where \( \xi_{-5} = (V_{\mathrm{W, \mathrm{local}}} / c)^2 \)/$10^{-5}$ is the acceleration parameter, \( B_{\mathrm{apex}} \) is the post-shock magnetic field strength at the apex of the WCR, \( R_{13} = R_{\mathrm{sh,W}} / 10^{13}~\mathrm{cm} \) is the WR shock standoff distance scaled to \(10^{13}~\mathrm{cm}\), \( v_3 = V_{\mathrm{W, local}} / 10^8~\mathrm{cm\,s^{-1}} \) is the wind speed in units of \(1000~\mathrm{km\,s^{-1}} \), and \( \dot{M}_{-4} = \dot{M}_{\mathrm{W}} / (10^{-4}~M_\odot\,\mathrm{yr}^{-1}) \) is the normalized mass-loss rate.

In the escape-limited regime, where the dominant limitation is the spatial confinement inside the WCR, the maximum energy is:
\begin{equation}
E_{p,\mathrm{max}} \approx 30\,\xi_{-5}\,B_{\mathrm{apex}}\,R_{13} / v_3 \quad \mathrm{TeV}.
\end{equation}

The choice between these regimes is made dynamically at each orbital phase by evaluating the critical condition:
\begin{equation}
\dot{M}_{-4} > 0.2\,R_{13}\,v_3^2.
\end{equation}
If this inequality is satisfied, the loss-limited formula is used; otherwise, the escape-limited one applies. This approach ensures that the maximum energy of protons responds sensitively to phase-dependent variations in magnetic field strength, wind velocity, and shock standoff distance.

Even after successful acceleration, relativistic protons may escape the WCR via diffusion before undergoing hadronic interactions. This escape reduces the hadronic gamma-ray yield. We account for this effect using the retention fraction, \( f_{\mathrm{ret}}(E) \), defined as the ratio of the diffusion timescale to the total timescale including both diffusion and hadronic interactions:
\begin{equation}
f_{\mathrm{ret}}(E) = \frac{\tau_{\mathrm{diff}}}{\tau_{\mathrm{diff}} + \tau_{\mathrm{loss}}}.
\end{equation}
The isotropic diffusion timescale is estimated as:
\begin{equation}
\tau_{\mathrm{diff}} = \frac{R_{\mathrm{sh,W}}^2}{6\,D(E)},
\end{equation}
The diffusion coefficient \( D(E) \) is modeled using a power-law energy scaling:
\begin{equation}
D(E) = D_0 \left( \frac{E}{1~\mathrm{MeV}} \right)^\delta,
\end{equation}
with \( \delta = 0.33 \), corresponding to Kolmogorov turbulence, and \( D_0 \sim 10^{19}~\mathrm{cm^2\,s^{-1}} \), representative of turbulent conditions in shocked stellar winds. The choice of diffusion coefficient is consistent with those explored in \cite{reitberger17}, and orders of magnitude less than that found in the interstellar medium (see, for e.g., \citealt{ptuskin06,putze10, genolini19, desarkar21}).
The hadronic interaction timescale is given by:
\begin{equation}
\tau_{\mathrm{loss}} = \frac{1}{n_{\mathrm{w, shocked}}\,\sigma\,c},
\end{equation}
where \( n_{\mathrm{w, shocked}} \) is the post-shock WR wind material number density, $\sigma \approx 3 \times 10^{-26}~\mathrm{cm^2}$ is the cross-section for hadronic interactions, and \( c \) is the speed of light. 
This retention factor modulates the injected proton spectrum, particularly at higher energies where escape becomes more probable. In denser or more extended shocks, \( f_{\mathrm{ret}} \to 1 \); in smaller or more diffusive regions, it drops significantly.

The fraction of the WR wind that impinges upon the WCR is determined by the wind momentum ratio $\eta_{\phi}$ given in Eq. \ref{eq1}, which controls the opening half angle of the conical shock surface. The shock-opening half angle \( \theta \) is approximated in Eq. \ref{eq2}. The fraction of the WR wind intercepted by the WCR is then computed via the solid angle:
\begin{equation}
f_{\mathrm{shock}} = \frac{2 \pi(1 - \cos\theta)}{4 \pi}.
\end{equation}
This factor is used to compute the fraction of wind energy diverted into the shock.

To account for the phase-dependent size of the acceleration region, we introduce a volume scaling factor:
\begin{equation}
f_{\mathrm{vol}} = \left( \frac{R_{\mathrm{sh,W}}}{R_{\mathrm{sep,peri}}} \right)^{\gamma_{\mathrm{vol}}},
\end{equation}
where \( R_{\mathrm{sh,W}} \) reflects the extent of the WCR on the WR side, and \( R_{\mathrm{sep,peri}} = a (1 - e) \) is the periastron separation used as a reference scale. A cubic exponent \( \gamma_{\mathrm{vol}} = 3 \) reflects the volumetric scaling of a conical or quasi-spherical acceleration region.
As the stellar separation increases toward apastron, the WCR expands, enlarging the acceleration volume and enhancing proton conversion. Near periastron, the WCR contracts, reducing both the acceleration volume and the particle residence time, thereby suppressing acceleration efficiency. By incorporating \( f_{\mathrm{vol}} \) alongside the conical interception fraction \( f_{\mathrm{shock}} \) and the retention efficiency \( f_{\mathrm{ret}}(E) \), we consistently capture the phase-dependent geometry and energetics of the system in the resulting hadronic gamma-ray emission from our simplified model.

The total kinetic power of the WR wind is given by:
\begin{equation}
L_{\mathrm{wind,W}} = \frac{1}{2} \dot{M}_{\mathrm{W}} V_{\mathrm{W, local}}^2.
\end{equation}
However, not all of this energy is available instantaneously for particle acceleration. The relevant timescale is the advection time:
\begin{equation}
t_{\mathrm{adv}} = \frac{3 R_{\mathrm{sh,W}}}{V_{\mathrm{W, local}}},
\end{equation}
accounting for the finite residence time of the particles in the acceleration zone. The total wind energy budget at each phase is then:
\begin{equation}
E_{\mathrm{wind}} = L_{\mathrm{wind,W}} \cdot t_{\mathrm{adv}}.
\end{equation}

The injection spectrum of relativistic protons is assumed to follow a power law in energy with an exponential cutoff at the maximum energy \( E_{p,\mathrm{max}} \),
\[
\frac{dN}{dE} = \mathcal{A} \, E^{-\alpha} \, \exp\left(-\frac{E}{E_{p,\mathrm{max}}}\right), 
\]
with spectral index \( \alpha \approx 2.2 \), as expected from standard DSA in shocks. To ensure that the total kinetic energy injected into protons is properly normalized, the normalization constant \( \mathcal{A} \) is computed as,

\begin{equation}
\mathcal{A} = \frac{\eta_p \cdot f_{\mathrm{shock}} \cdot f_{\mathrm{vol}} \cdot f_{\mathrm{ret}}(E) \cdot E_{\mathrm{wind}}}{\int E^{- \alpha + 1} \, \exp\left(-\frac{E}{E_{p,\mathrm{max}}}\right) \ dE},
\end{equation}
where \( \eta_p \) is the proton acceleration efficiency (typically \(  10^{-4} - 10^{-1} \)). Finally, the total energy injected into protons at a given orbital phase is obtained by integrating:
\begin{equation}
E_{p,\mathrm{inj}} = \int E \ \frac{dN}{dE} \, dE.
\end{equation}

This formalism yields a proton injection energy that is naturally modulated by orbital phase, with higher values near apastron (due to increased WCR volume and residence time) and suppressed values near periastron (due to compression and escape). 
Notably, the effective kinetic power available for proton acceleration in our model is governed by the product \( f_{\mathrm{shock}} \cdot f_{\mathrm{vol}} \cdot f_{\mathrm{ret}}(E) \), which encapsulates the geometric, volumetric, and confinement-related modulation of the injection process. This formulation closely resembles the effective injection factor \( f_{\mathrm{eff}} \) introduced in \citet{debecker13}. While \( f_{\mathrm{eff}} \) is defined in a system-specific context, in our simplified model it is effectively realized through this phase-dependent product, providing a physically motivated analogue that captures the evolving energetics of the wind interaction zone.

Finally, the phase-dependent hadronic $\gamma$-ray emission from $\gamma^2$~Velorum is computed by convolving the spectrum of injected relativistic protons with the local post-shock WR wind density, which serves as the target for hadronic interactions. At each orbital phase, the proton injection spectrum is folded with the shocked WR wind target density \( n_{\mathrm{w, shocked}} \), to calculate the $\pi^0$-decay $\gamma$-ray emissivity, while taking into account the nuclear enhancement factor $\epsilon_{\mathrm{enh}}$. 
The resulting emission reflects both the energy distribution of the accelerated protons and the evolving ambient density conditions across the orbit. These calculations are performed using the \texttt{GAMERA} library \citep{hahn15, hahn22}, which implements the hadronic interaction formalism and pion production cross-sections as parameterized in \citet{kafe14}.

{ 
As discussed earlier, the O-star shock remains in the adiabatic regime throughout all orbital phases. Nevertheless, the higher unbraked velocity of the O-star wind, combined with its comparatively lower mass-loss rate, leads to a shorter advection timescale, reduced kinetic energy available for particle acceleration, and a lower post-shock density for hadronic interactions. Consequently, the O-star wind contributes negligibly to the overall gamma-ray emission when compared to the denser and more energetically dominant WR wind.

}

\subsection{$\gamma$--$\gamma$ absorption}\label{gamma}

High-energy gamma rays traversing the intense radiation fields of massive stars in the CWB system can be absorbed via pair production process ($\gamma + \gamma \rightarrow e^+ + e^-$), resulting in orbital-phase-dependent attenuation. In $\gamma^2$~Velorum, we compute the total $\gamma$--$\gamma$ optical depth $\tau_{\gamma\gamma}$ by considering the finite sizes of both stars, their anisotropic blackbody photon fields, orbital inclination, and the location of the emission region at the WCR.

Both stars are treated as finite-sized blackbody emitters. The spectral number density of target photons at a given energy $\epsilon$ is described by the Planck distribution:
\begin{equation}
n(\epsilon) = \frac{2}{h^3 c^3} \cdot \frac{\epsilon^2}{\exp(\epsilon / k_B T_*) - 1},
\end{equation}
where $T_*$ is the effective temperature of the star, and $h$, $c$, and $k_B$ are the usual physical constants.

The $\gamma$--$\gamma$ pair production cross-section is calculated using the Breit–Wheeler formula, which depends on the interaction angle $\psi$ between the gamma-ray and target photon \citep{breit34, gould67}:
\begin{equation}
s = \frac{E_\gamma \epsilon (1 - \cos\psi)}{2 (m_e c^2)^2},
\end{equation}
\begin{equation}
\sigma_{\gamma\gamma}(s) = \frac{3\sigma_T}{16} (1 - \beta^2) \left[ (3 - \beta^4) \ln \left( \frac{1 + \beta}{1 - \beta} \right) - 2 \beta (2 - \beta^2) \right],
\end{equation}
where $\beta = \sqrt{1 - 1/s}$, and $\sigma_T$ is the Thomson cross-section. The cross-section is non-zero only above the pair production threshold ($s \geq 1$).

The optical depth is computed by integrating along the gamma-ray path starting from the shock apex and extending outward along the line of sight. At each integration step, the angle $\psi$ and distance $d$ from the point to the star are used to compute the solid angle subtended by the stellar disk:
\begin{equation}
\Omega_* = 2\pi \left( 1 - \sqrt{1 - \left( \frac{R_*}{d} \right)^2 } \right),
\end{equation}
and the minimum target photon energy for pair production is:
\begin{equation}
\epsilon_{\min} = \frac{2 (m_e c^2)^2}{E_\gamma (1 - \cos\psi)}.
\end{equation}
The differential optical depth along the line of sight over the path length $dz$ is given by,
\begin{equation}
\frac{d\tau_{\gamma\gamma}}{dz} = \frac{1 - \cos\psi}{4\pi} \, \Omega_* \int_{\epsilon_{\min}}^\infty n(\epsilon) \sigma_{\gamma\gamma}(E_\gamma, \epsilon, \psi) \, d\epsilon,
\end{equation}
which is then integrated over the path to obtain the total optical depth $\tau_{\gamma\gamma}$. If the gamma-ray path intersects a stellar surface ($d < R_*$), we set $\tau_{\gamma\gamma} \gg 1$ to simulate full absorption. Note that the geometry of the stars and the interaction region are projected into the observer's frame by applying an inclination angle rotation about the orbital plane.

The total optical depth is the sum of contributions from both stars:
\begin{equation}
\tau_{\gamma\gamma}(E_\gamma) = \tau_{\rm \gamma\gamma, O}(E_\gamma) + \tau_{\rm \gamma\gamma, W}(E_\gamma),
\end{equation}
where $\tau_{\rm \gamma\gamma, O}$ and $\tau_{\rm \gamma\gamma, W}$ are the individual optical depths from the O- and WR star photon fields, respectively. The final attenuation factor is given by $\exp(-\tau_{\gamma\gamma})$, and varies with orbital phase and gamma-ray energy for a constant inclination angle. This treatment captures the anisotropic and phase-dependent absorption relevant for modeling the observed modulation of high-energy emission from $\gamma^2$~Velorum.


\section{Results}\label{results}

\begin{figure*}
    \centering
    \includegraphics[width=\columnwidth]{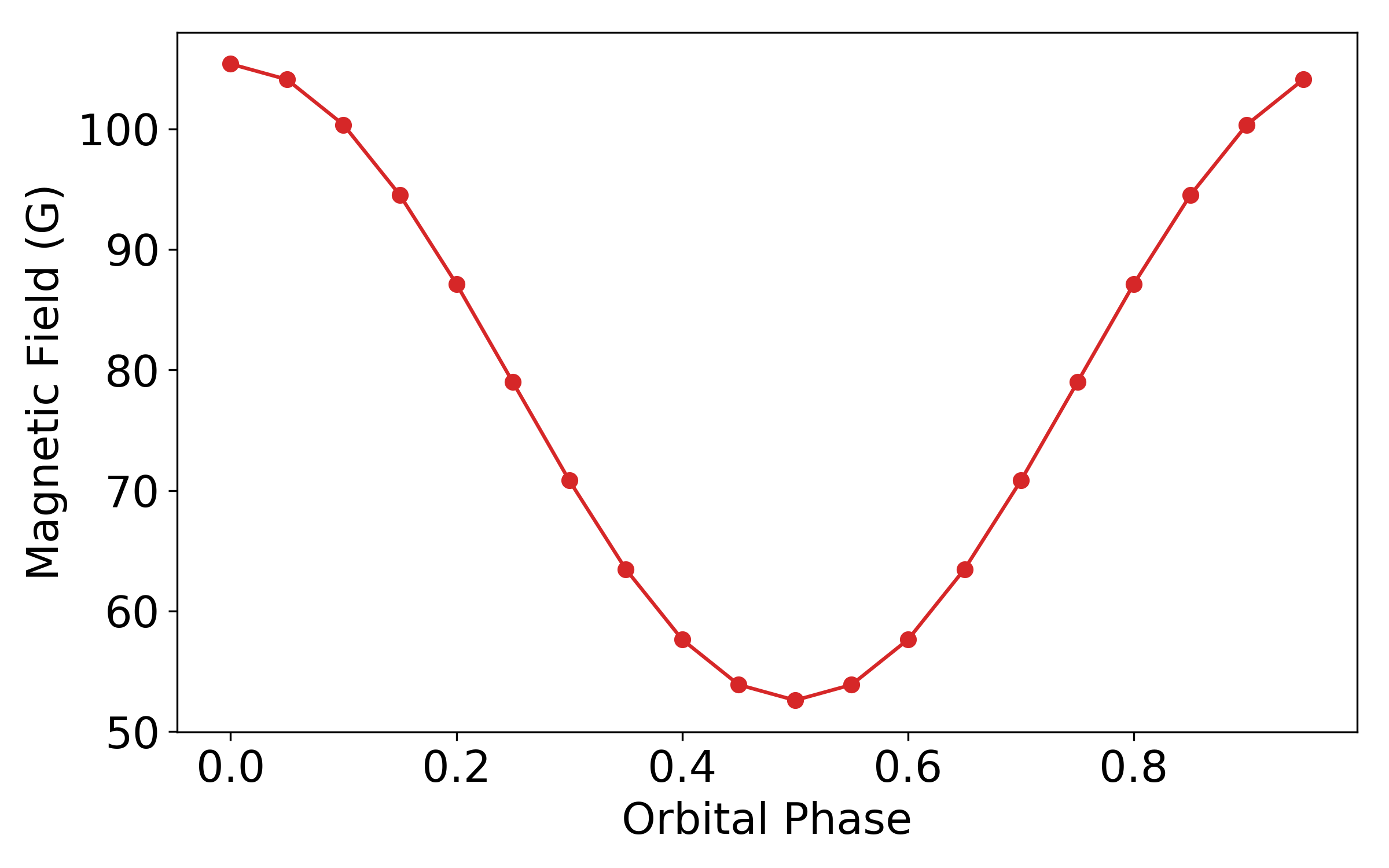}
    \includegraphics[width=\columnwidth]{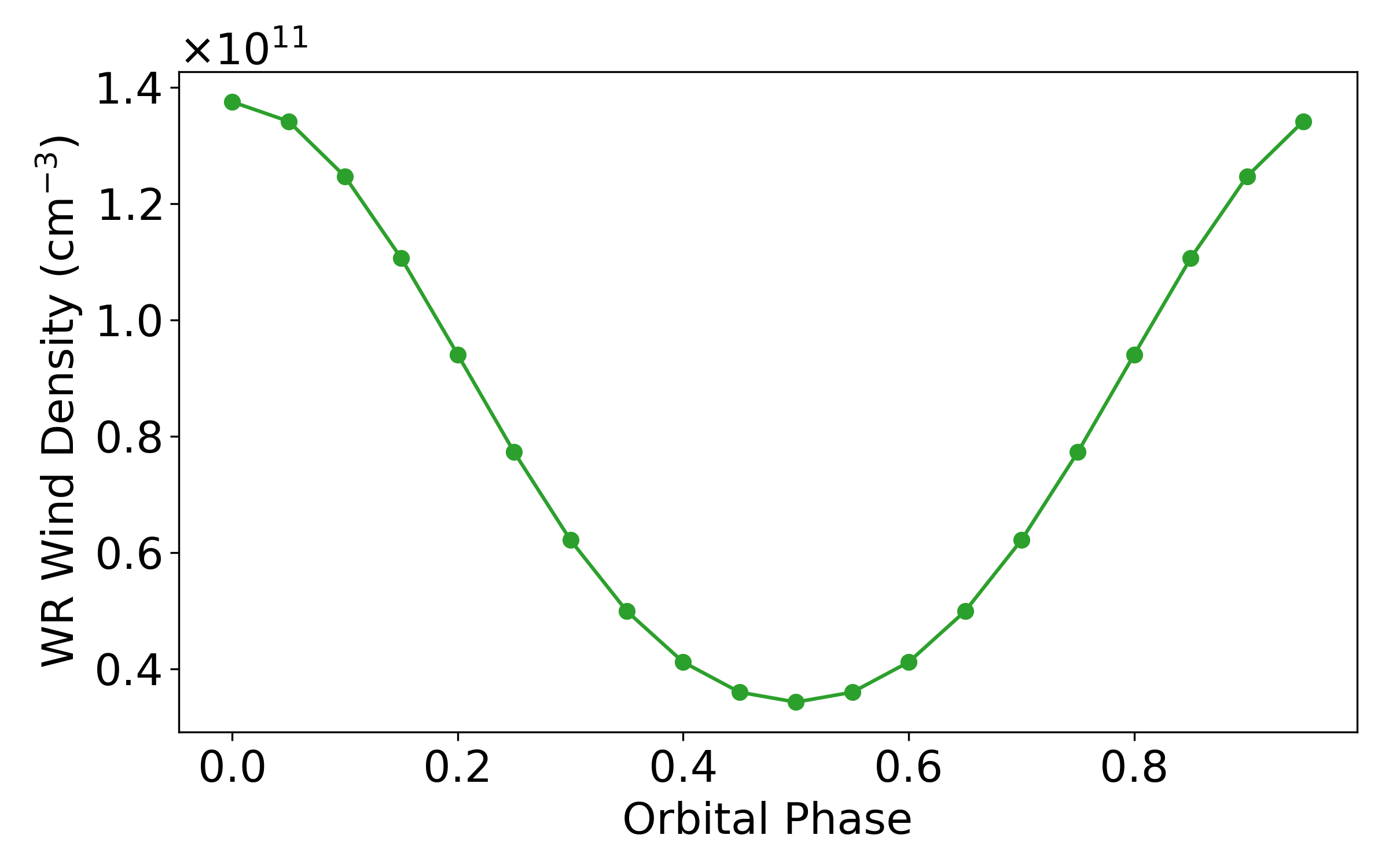}
    \includegraphics[width=\columnwidth]{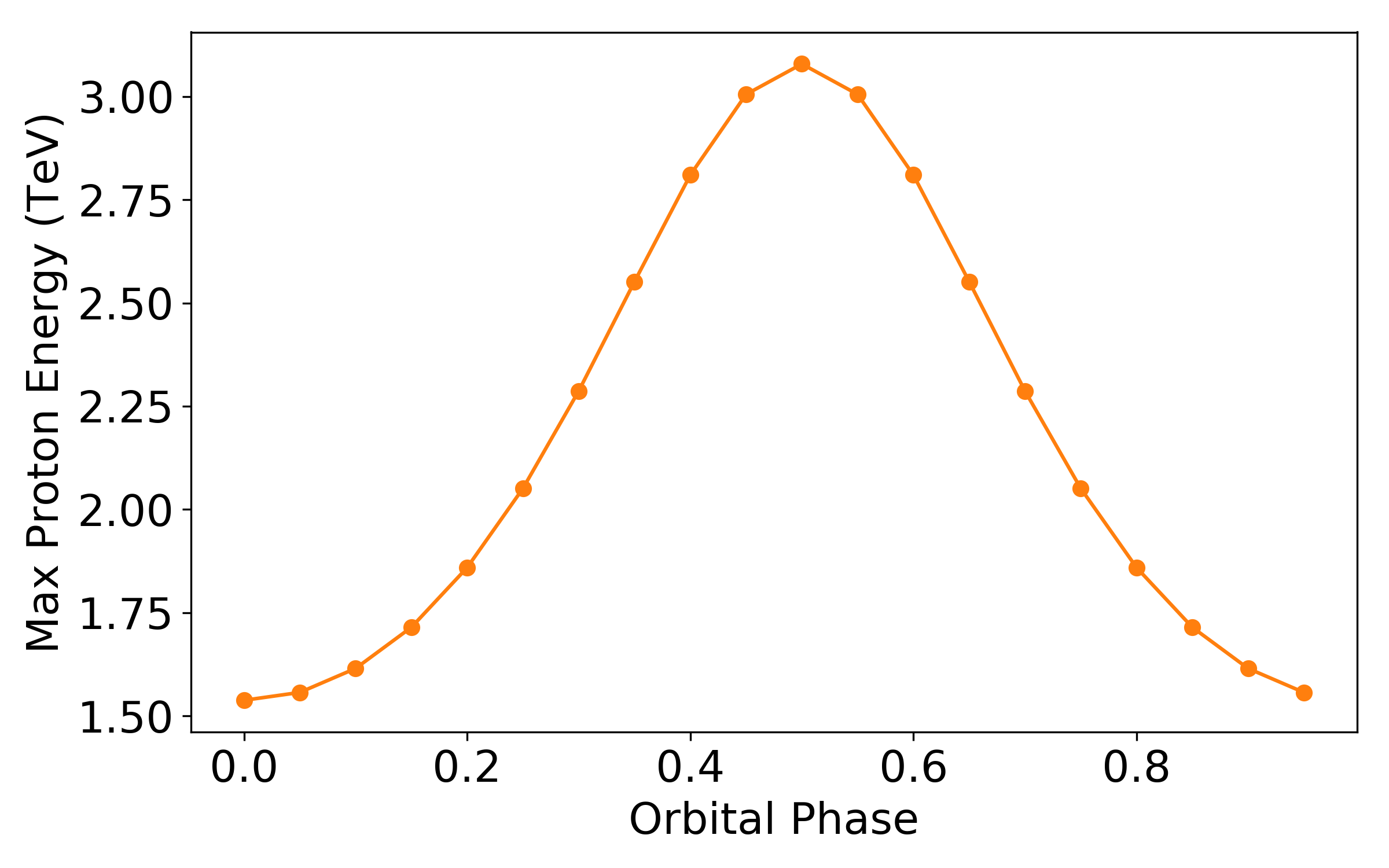}
    \includegraphics[width=\columnwidth]{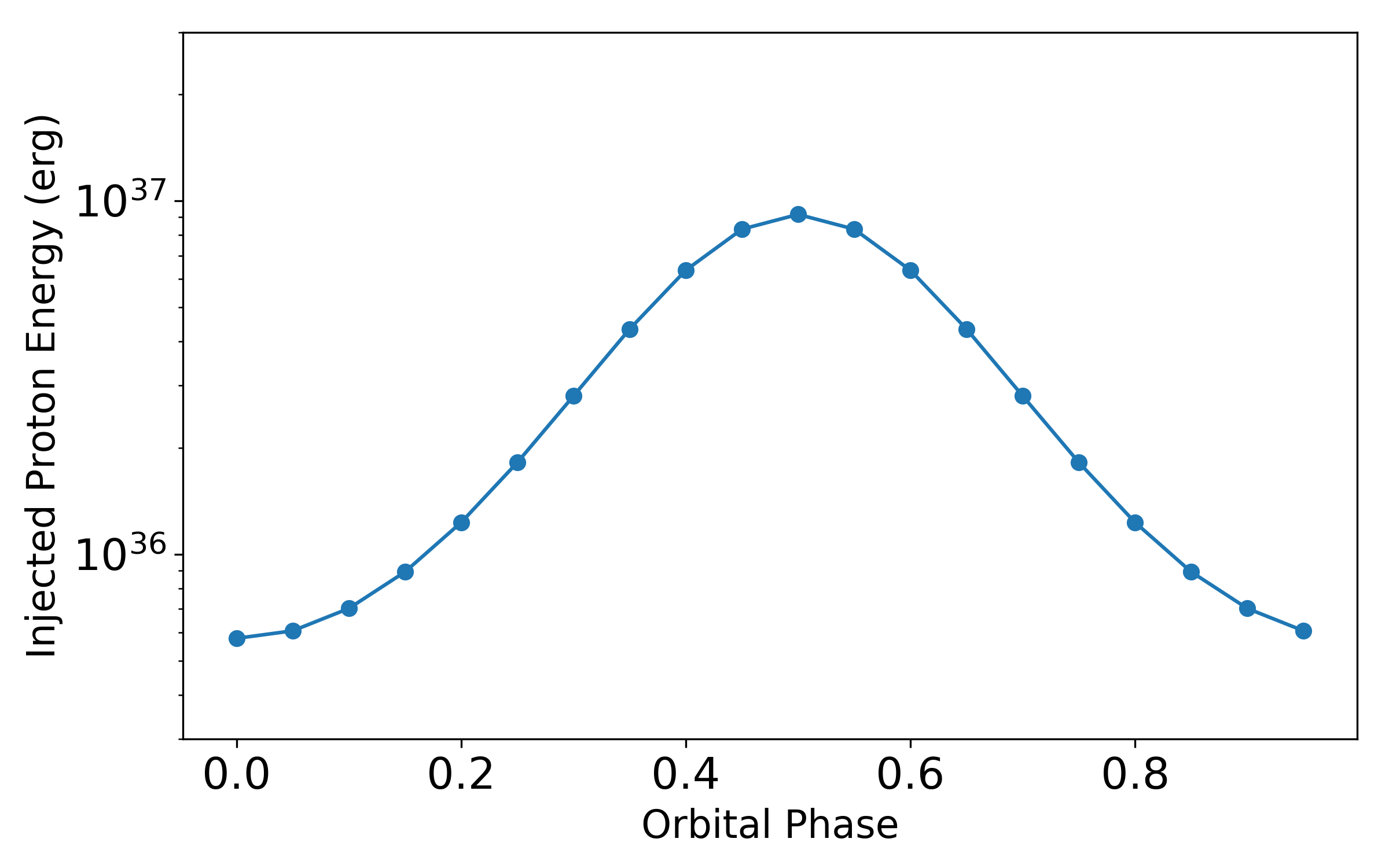}
    \caption{Variation of the magnetic field (top left), post-shock WR wind number density (top right), proton maximum energy (bottom left), and injected proton energy budget (bottom right) with orbital phases are shown in this plot.}
    \label{fig1}
\end{figure*}

In this section, we present the results of our phase-dependent hadronic emission model for $\gamma^2$~Velorum, which captures the evolving interplay between stellar wind geometry, proton acceleration, and $\gamma$--$\gamma$ absorption across the binary orbit. Our simulations sample 20 evenly spaced orbital phases and account for phase-dependent variations in the WCR magnetic field strength, the shocked WR wind density, and the total energy injected into relativistic protons.

The key physical conditions in the acceleration region---the magnetic field strength, post-shock WR wind density, and maximum attainable proton energy---exhibit systematic variations across orbital phase due to the evolving geometry of the WCR. The magnetic field at the WCR apex increases toward periastron due to enhanced compression of the WR wind.
Similarly, the post-shock WR wind density reaches its peak near periastron as the standoff distance decreases, leading to a denser target for hadronic interactions. 
However, the efficiency of gamma-ray production is suppressed in this phase because the compact WCR limits both the spatial volume available for particle acceleration and the retention of high-energy protons. 
The maximum energy of accelerated protons also varies with phase, with higher values near apastron, where larger WCR volume and longer escape timescales facilitate more efficient acceleration. Note that the maximum proton energy calculated from our model comes out just above $\sim$ 1 TeV, which is consistent with that calculated in \cite{reitberger17}.
At each orbital phase, the total energy injected into relativistic protons is computed by combining the WR wind kinetic energy with the conical shock interception fraction, volume scaling factor, and retention efficiency. Along with the available WR wind energy budget $E_{\mathrm{wind}}$, the injection energy in protons \( E_{p,\mathrm{inj}} \) exhibits a clear enhancement near apastron (\( \phi \approx 0.5 \)). This increase in \( E_{p,\mathrm{inj}} \) arises primarily from the geometric expansion of the WCR, which enhances proton conversion by increasing the acceleration volume, extending the diffusion timescale, and increasing the probability of proton interaction before escape. In contrast, the periastron phase (\( \phi \approx 0 \)) is characterized by a compact, dense region with reduced residence time and acceleration volume, leading to a suppression of the injected proton energy budget and hadronic gamma-ray emission.
The orbital variation of these parameters is presented in Fig. \ref{fig1} of the paper.

\begin{figure*}
    \centering
    \includegraphics[width=\columnwidth]{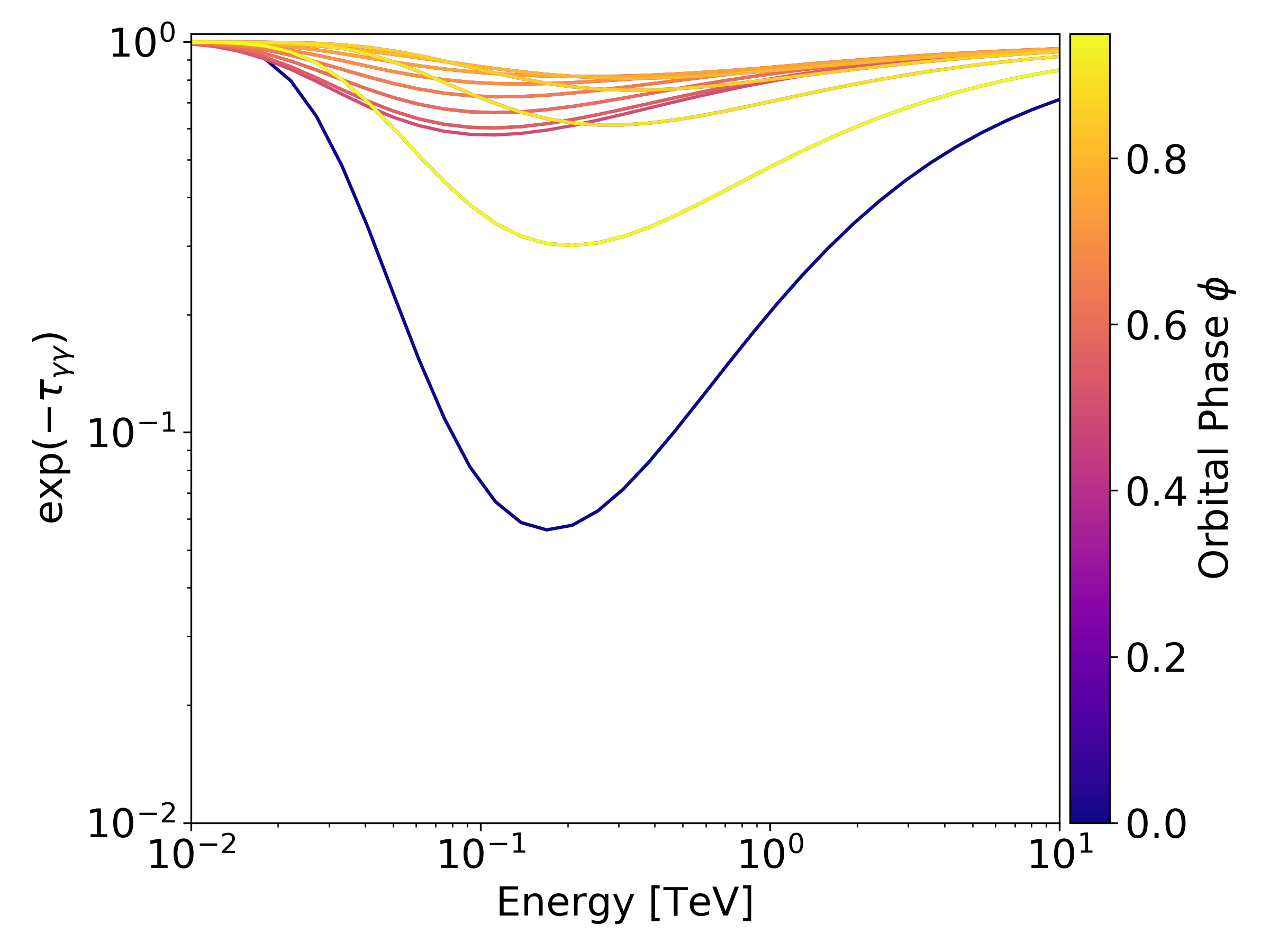}
    \includegraphics[width=\columnwidth]{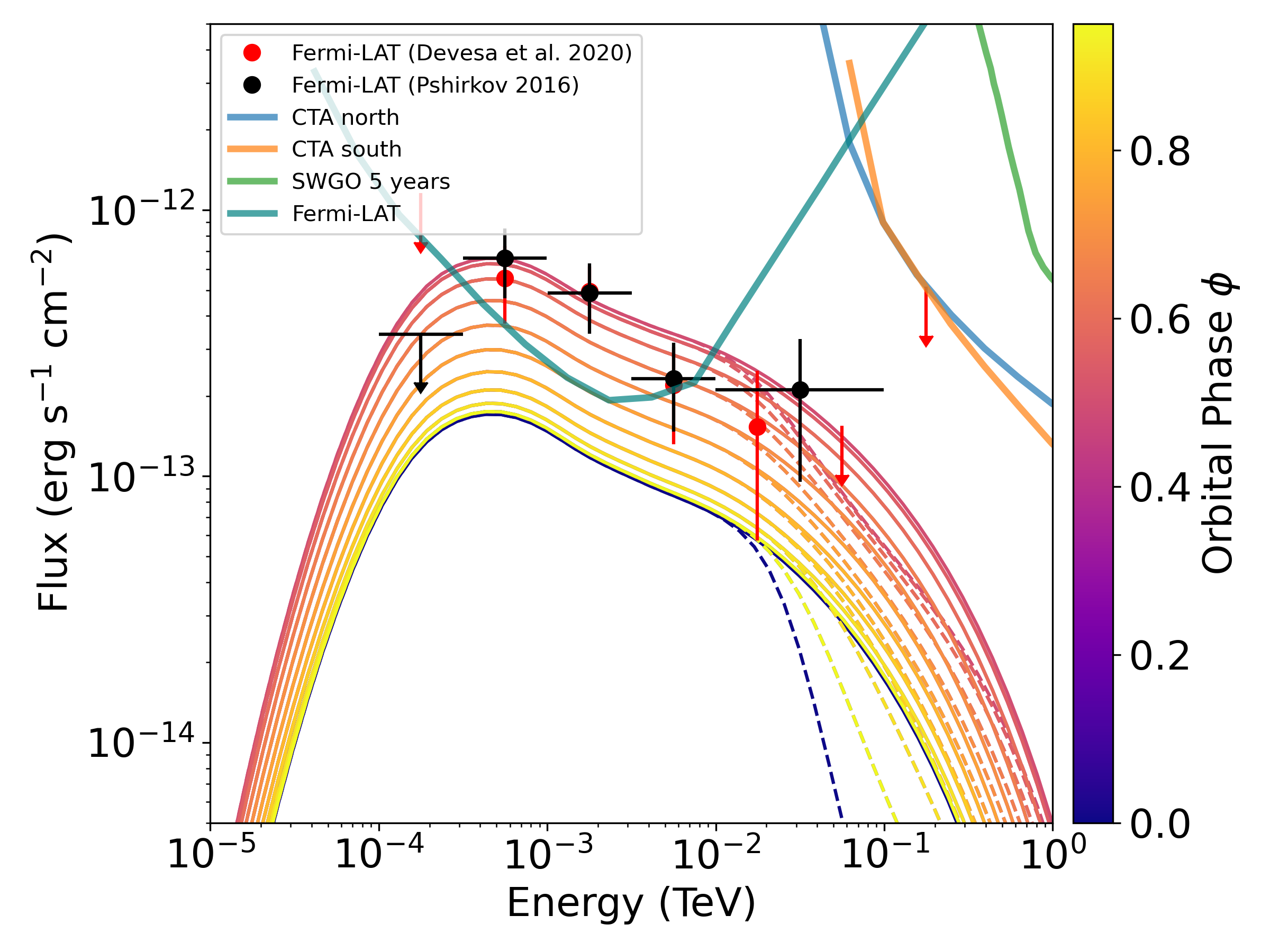}
    \caption{The left panel shows the attenuation factor \( \exp(-\tau_{\gamma\gamma}) \) as a function of energy and orbital phase used in this model. The right panel shows the phase-dependent model SEDs varying across the orbital phases, along with the {\it Fermi}-LAT data points \citep{pshirkov16, devesa20}. The CTA, SWGO, and {\it Fermi}-LAT sensitivities are also shown in the figure for comparison purposes. The {\it Fermi}-LAT sensitivity corresponds to the 10-year benchmark (P8R3\_SOURCE\_V3), assuming a point source with a power-law spectrum in a uniform diffuse background. It serves as a reference only and does not represent a strict detection threshold. Variations in background intensity and exposure at the location of $\gamma^2$~Velorum, along with low test statistics (TS) or large uncertainties, can cause some data points to appear below this curve. The solid lines represent the unabsorbed SEDs, whereas the dashed lines signify the SEDs affected by $\gamma-\gamma$ absorption.}
    \label{fig2}
\end{figure*}

Using the computed proton injection spectra at each orbital phase, we evaluate the resulting hadronic SEDs, both with and without $\gamma$--$\gamma$ absorption. The required proton acceleration efficiency to match the observed {\it Fermi}-LAT SED is found to be \( \eta_p \sim 10^{-4} \), which is somewhat lower than the value adopted in \citet{reitberger17}, yet remains within a physically plausible range. We note that \( \eta_p \) is sensitive to the choice of the diffusion normalization \( D_0 \) (\(\sim 10^{19}~\mathrm{cm^2\,s^{-1}} \) in this paper); larger \( D_0 \) values lead to lower retention efficiency and thus require higher \( \eta_p \) to compensate. However, in the absence of tighter observational constraints, a broader parameter exploration is not currently feasible.
The $\gamma$--$\gamma$ absorption significantly suppresses the high-energy flux, particularly at periastron, where the photon field density is highest due to the smaller binary separation. This attenuation is illustrated in the left panel of Figure~\ref{fig2}, which shows the energy-dependent transmission factor \( \exp(-\tau_{\gamma\gamma}) \) across orbital phase.
The right panel of Figure~\ref{fig2} compares the resulting phase-dependent SEDs to {\it Fermi}-LAT observations from \citet{devesa20} and \citet{pshirkov16}. Our model, with and without $\gamma$--$\gamma$ absorption, reproduces both the spectral shape and flux normalization reasonably well. While \citet{devesa20} interpreted the {\it Fermi}-LAT spectrum as arising either entirely from apastron or phase-averaged emission, we assume that the observed $\gamma$-ray flux originates predominantly near apastron. In this scenario, the suppressed emission at periastron falls below the LAT sensitivity threshold, consistent with its non-detection during those phases.
Finally, given the limited maximum energy of accelerated protons (\( E_{p,\mathrm{max}} \sim \mathrm{a \ few\ TeVs} \)), our model shows a steep cutoff in the gamma-ray SED above $\sim$ 100 GeV, suggesting that $\gamma^2$~Velorum is unlikely to be detected by next-generation TeV instruments such as the Cherenkov Telescope Array (CTA) or the Southern Wide-field Gamma-ray Observatory (SWGO).

Next, we compute the phase-dependent gamma-ray flux integrated in energy between 100~MeV and 500~GeV across the orbit of $\gamma^2$~Velorum. The full phase-energy-integrated flux, accounting for $\gamma$--$\gamma$ absorption, is found to be $\sim 8.6 \times 10^{-7}~\mathrm{MeV~cm^{-2}~s^{-1}}$, corresponding to a photon flux of $\sim 1.5 \times 10^{-9}~\mathrm{ph~cm^{-2}~s^{-1}}$. The values are a few factors lower than those reported in \citet{devesa20}, with the discrepancy likely arising from differences in the assumed proton spectral index, as our diffusion-limited model yields a harder slope compared to their softer assumption. To assess flux modulation across the orbit, we adopt the phase ranges $\phi = 0.25$--$0.75$ (apastron) and $\phi = 0.0$--$0.25$ and $0.75$--$1.0$ (periastron), following \citet{devesa20}. The resulting apastron-to-periastron flux ratio of $\sim 2.4$ aligns well with the non-gated ($\sim 2.3$) and gated ($\sim 2.8$) values from that study, indicating that our model successfully reproduces the observed orbital modulation, which is dominated by WCR geometry, with absorption playing a secondary role.

\begin{figure}
    \centering
    \includegraphics[width=\columnwidth]{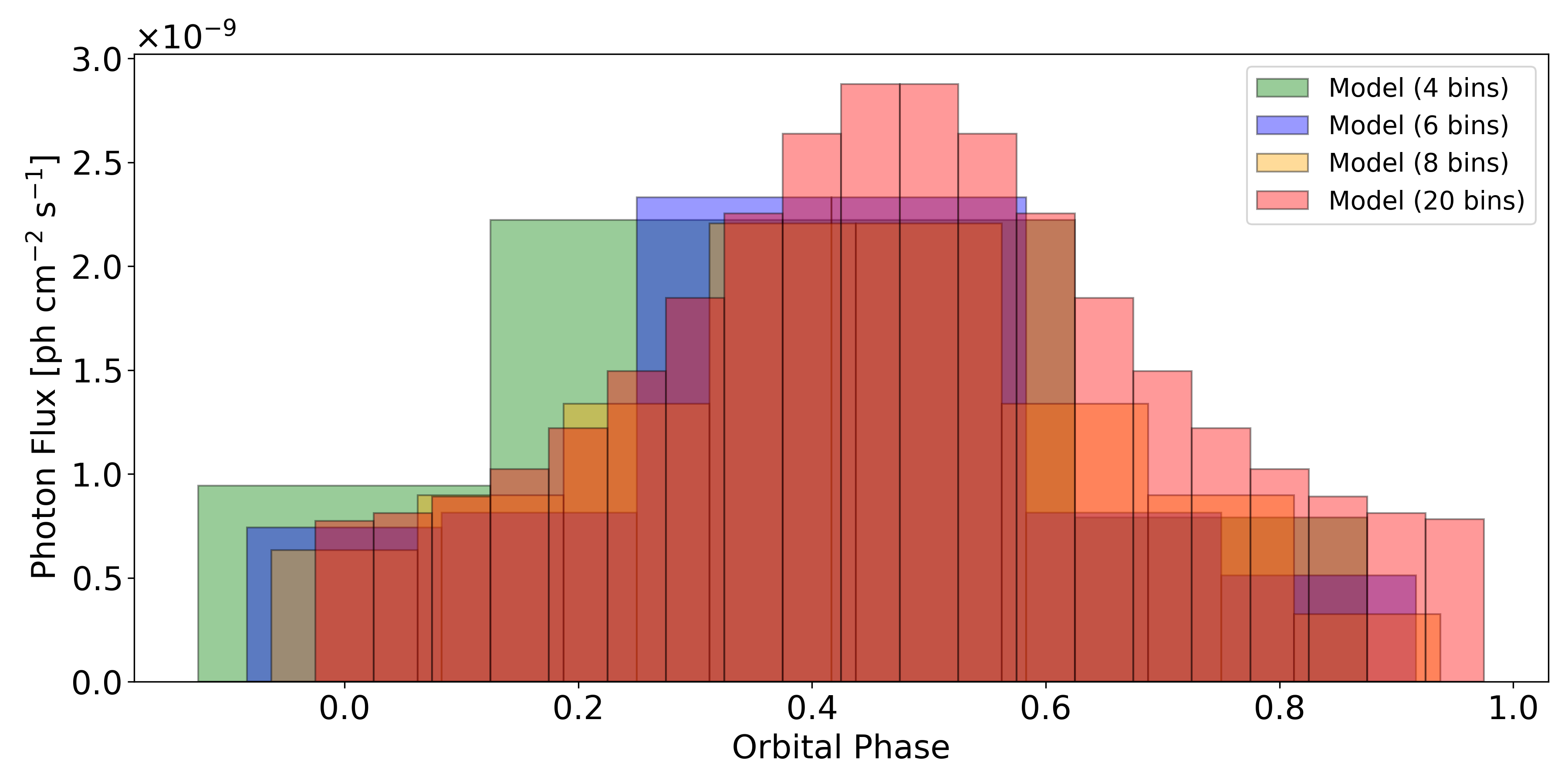}
    \caption{The variation of average photon flux per bin with orbital phase for a single orbit is shown in this figure. The orbit is divided into 4 bins (green), 6 bins (blue), 8 bins (yellow), and 20 bins (red). As usual, the periastron occurs at phase $\phi = 0$, and the apastron occurs at $\phi = 0.5$.}
    \label{fig3}
\end{figure}

To visually illustrate the orbital modulation, we construct phase-binned light curves using the photon flux integrated between 100~MeV and 500~GeV. The orbital phase interval \([0,1]\) is divided into \(N = 4\), 6, 8, and 20 equal bins. For each bin, we integrate the energy-integrated photon flux over the corresponding phase interval and normalize the result by the bin width, yielding the average photon flux per bin. These values are then used to generate the model gamma-ray light curves shown in Figure~\ref{fig3}. The resulting profiles clearly exhibit a flux minimum near periastron (\(\phi \approx 0\)) and a maximum near apastron (\(\phi \approx 0.5\)), in agreement with the observed modulation reported by \citet{devesa20}. While we do not attempt to fit the light curve data due to limited observational statistics, this exercise demonstrates that incorporating orbital-phase-dependent variations in WCR geometry is essential to reproduce the apastron/periastron flux contrast observed in \textit{Fermi}-LAT data.

{The phase-dependent behavior of gamma-ray emission contrasts sharply between $\gamma^2$~Velorum and $\eta$~Carinae, and can be understood in terms of both wind properties and shock cooling regimes. In the moderately eccentric system $\gamma^2$~Velorum, the relatively low mass-loss rate leads to longer hadronic interaction timescales, making diffusion a significant factor compared to the losses. The WR shock, in this case, is radiative, resulting in moderately enhanced post-shock compression and target densities. However, near periastron, the compact WCR limits the available acceleration volume and reduces the residence time of relativistic protons, lowering the efficiency of wind energy conversion into protons, thus leading to a suppresion of hadronic gamma-ray emission. At apastron, the expanded WCR increases both the acceleration volume and the retention efficiency, enhancing the probability of hadronic interactions and boosting the gamma-ray output.
In contrast, the highly eccentric system $\eta$~Carinae exhibits peak gamma-ray emission near periastron due to its extreme mass-loss rate, which generates a highly dense post-shock region. This condition significantly shorten the hadronic interaction timescale, making losses dominant over diffusion and enabling efficient emission despite the compactness of the WCR. Furthermore, the shock is expected to be much more strongly radiative in $\eta$~Carinae compared to $\gamma^2$~Velorum, thus producing higher compression ratio and elevated post-shock density that further amplify the gamma-ray yield. Although it will also make the shock structure more complicated, which in turn could impact the acceleration process. At apastron, the increased orbital separation reduces wind densities and interaction probability, leading to suppressed emission.
Thus, the key physical difference between the two systems lies not only in orbital geometry and eccentricity, but also in the prevailing shock cooling regimes and their influence on particle acceleration and target densities throughout the orbit.

}

Leptonic processes are not included in our model, as the strong stellar radiation field in $\gamma^2$~Velorum imposes severe inverse-Compton losses that limit electron energies to $\sim$10~MeV at the WCR apex and $\sim$100~MeV in the outer regions \citep{reitberger17}, making their contribution to GeV emission negligible. Thus, the high-energy output is expected to be hadron-dominated. Although the absorbed non-thermal radio component suggests the presence of accelerated electrons \citep{purton82, chapman99, debecker13}, we focus on the orbital modulation of the GeV emission, where leptonic contributions are likely subdominant.



\section{Concluding remarks}\label{conclusion}

We have developed a physically motivated, phase-dependent model of hadronic gamma-ray emission in the colliding-wind binary $\gamma^2$~Velorum, incorporating variations in WCR geometry, proton acceleration and escape, and $\gamma$--$\gamma$ absorption. This framework successfully reproduces the observed SED and explains the hints of orbital variability seen in the {\it Fermi}-LAT data.

Our results indicate that the primary driver of the flux variability is the evolving geometry of the WCR with orbital phase. The widening of the WCR at apastron enhances the acceleration volume and proton retention, increasing the efficiency of hadronic interactions and boosting gamma-ray output, and vice versa during periastron. While $\gamma$--$\gamma$ attenuation contributes to emission suppression near periastron, it alone cannot explain the apastron flux enhancement, underscoring the importance of intrinsic modulation.

Although based on a simple semianalytic treatment, our model demonstrates the key role of geometric effects in shaping high-energy emission from CWBs. Future efforts involving 3D MHD simulations will be essential to capture additional complexities—such as shock curvature, instabilities, magnetic field evolution, and dynamic WCR disruption—that can further influence particle acceleration and gamma-ray production. The present study provides a foundation for such advanced modeling and emphasizes the relevance of orbital geometry in interpreting the high-energy behavior of eccentric binary systems like $\gamma^2$~Velorum.


\begin{acknowledgements}
A.D.S. thanks the anonymous reviewer for helpful suggestions and constructive criticism, which vastly improved the quality of this work. A.D.S. is supported by the Juan de la Cierva JDC2023-052168-I grant, funded by MICIU/AEI/10.13039/501100011033 and the FSE+. This work is also supported by the grant PID2021-124581OB-I00 of MCIU/AEI/10.13039/501100011033 and 2021SGR00426 and the program Unidad de Excelencia María de Maeztu CEX2020-001058-M, and by MCIU with funding from European Union NextGeneration EU (PRTR-C17.I1).
\end{acknowledgements}

%
   \bibliographystyle{aa} 
   \bibliography{aa.bib} 
%

\end{document}